\pdfoutput=1

\documentclass[letterpaper,twocolumn,10pt]{article}

\newcommand{\asplossubmissionnumber}{125}

\usepackage{tikz}
\usepackage{tikzsymbols}
\usetikzlibrary{calc, positioning, arrows.meta, shapes.geometric, decorations,
  patterns, backgrounds, fit, patterns.meta, chains, scopes, shadows.blur}
\usepackage{pgf-umlsd}

\usepackage[sort,numbers]{natbib}
\usepackage{doi}
\usepackage{numprint}

\usepackage{subcaption}

\usepackage{amsmath,amssymb,latexsym}
\usepackage[binary-units=true]{siunitx}

\usepackage{listings}
\usepackage{newfloat} 
\DeclareFloatingEnvironment[%
 fileext=lool,
 name=Listing
]{listing}

\usepackage[%
 capitalise,  
 nameinlink,  
 noabbrev,    
] {cleveref}

\usepackage{paralist}

\usepackage{todonotes}
\usepackage{csquotes}
\usepackage{pifont}
\usepackage{layouts}
\usepackage{booktabs}

\makeatletter
\renewcommand{\paragraph}{%
  \@startsection{paragraph}{4}%
  {\z@}{.3ex \@plus .2ex \@minus .2ex}{-1em}%
  {\normalfont\normalsize\bfseries}%
}
\makeatother

\newcommand*{\infiniband}{InfiniBand}
\newcommand*{\Infiniband}{\infiniband}
\newcommand*{\ibverbs}{IB~verbs}
\newcommand*{\nakstopped}{\texttt{NAK\_STOPPED}}
\newcommand*{\Spaused}{\emph{Paused}}
\newcommand*{\Sstopped}{\emph{Stopped}}
\newcommand*{\sendbw}{\texttt{ib\_send\_bw}}
\newcommand*{\perftest}{\texttt{perftest}~\cite{perftest}}

\newcommand{\name}{MigrOS}
\begin{document}

\date{}

\title{\name{}: Transparent Operating Systems Live Migration Support for
  Containerised RDMA-applications}

\author{
{\rm Maksym Planeta}\\
TU Dresden
\and
{\rm Jan Bierbaum}\\
TU Dresden
\and
{\rm Leo Sahaya Daphne Antony}\\
AMOLF
\and
{\rm Torsten Hoefler}\\
ETH Z\"urich
\and
{\rm Hermann H\"artig}\\
TU Dresden
} 

\maketitle

\begin{abstract}
  Major data centre providers are introducing RDMA-based networks for their
  tenants, as well as for operating their underlying infrastructure.
  In comparison to traditional socket-based network stacks, RDMA-based networks
  offer higher throughput, lower latency, and reduced CPU overhead.
  However, RDMA networks make transparent checkpoint and migration operations
  much more difficult.
  The difficulties arise because RDMA network architectures remove the OS from
  the critical path of communication.
  As a result, the OS loses control over active RDMA network connections, required
  for live migration of RDMA-applications.
  This paper presents \name{}, an OS-level architecture for transparent live
  migration of RDMA-applications.
  \name{} offers changes at the OS software level and small changes to the RDMA
  communication protocol.
  As a proof of concept, we integrate the proposed changes into SoftRoCE, an
  open-source kernel-level implementation of an RDMA communication protocol.
  We designed these changes to introduce no runtime overhead, apart from the
  actual migration costs.
  \name{} allows seamless live migration of applications in data centre
  settings.
  It also allows HPC clusters to explore new scheduling strategies, which
  currently do not consider migration as an option to reallocate the resources.
\end{abstract}



\section{Introduction}

Cloud computing is undergoing a phase of rapidly increasing
network performance.
This trend implies higher requirements on the data and packet processing rate
and results in the adoption of high-performance network stacks~\cite{ibta, dpdk,
  arrakis, ix, mellanox-vma, vsocket, slimos}.
RDMA network architectures address this demand by offloading packet processing
onto specialised circuitry of the network interface controllers~(RDMA NICs).
%
These RDMA NICs process packets much faster than CPUs.
User applications communicate directly with the NICs to send and receive
messages using specialised RDMA-APIs, like \ibverbs{}\footnote{\ibverbs{} is the
  most common low-level API for RDMA networks.}.%
%
%
This direct access minimises network latency, which made RDMA networks
ubiquitous in HPC~\cite{hessGROMACSAlgorithmsHighly2008,
  kaleCHARMPortableConcurrent1993, liuHighPerformanceRDMAbased2003, ucx} and
increasingly more accustomed in the data centre
context~\cite{gaoNetworkRequirementsResource2016, pokeDAREHighPerformanceState2015,
  mitchellUsingOneSidedRDMA2013}.
As a result, major data centre providers already offer RDMA connectivity for the
end-users~\cite{ElasticFabricAdapter,HighPerformanceComputing2020}.

Similarly, containers have also become ubiquitous for lightweight virtualisation
in data centre settings.
Containerised applications do not depend on the software stack of the host, thus
greatly simplifying distributed application deployment and administration.
However, RDMA networks and containerisation come at odds, when employed
together: The former try to bring applications and underlying hardware
\enquote{closer} to each other, whereas the latter facilitates the opposite.
This paper, in particular, addresses the issue of migratability of containerised
RDMA-applications through OS-level techniques.

%

\newcommand{\basewidth}{0.78cm}
\newcommand{\swheight}{0.78cm}
\newcommand{\baseskip}{1pt}

\tikzdeclarepattern{
  name=lines,
  parameters={
    \pgfkeysvalueof{/pgf/pattern keys/size},
    \pgfkeysvalueof{/pgf/pattern keys/angle},
    \pgfkeysvalueof{/pgf/pattern keys/line width},
  },
  bounding box={(-.1pt,-.1pt) and
    (\pgfkeysvalueof{/pgf/pattern keys/size}+.1pt,
    \pgfkeysvalueof{/pgf/pattern keys/size}+.1pt)},
  tile size={(\pgfkeysvalueof{/pgf/pattern keys/size},
    \pgfkeysvalueof{/pgf/pattern keys/size})},
  tile transformation={rotate=\pgfkeysvalueof{/pgf/pattern keys/angle}},
  defaults={
    size/.initial=5pt,
    angle/.initial=0,
    line width/.initial=.4pt,
  },
  code={
    \draw[line width=\pgfkeysvalueof{/pgf/pattern keys/line width}]
    (0,0) -- (\pgfkeysvalueof{/pgf/pattern keys/size},
    \pgfkeysvalueof{/pgf/pattern keys/size});
  }
}

\tikzset{
  sw/.style = {
    node distance = 0pt
  },
  kernel/.style = {
    draw,
    rectangle,
    minimum width = (\basewidth - 0.1cm),
    minimum height = \swheight,
    node distance = \baseskip,
    label={[anchor=base]above,label distance=0.7ex:{\emph{\small Kern}}},
  },
  app/.style = {
    draw,
    rectangle,
    minimum width = (\basewidth + 0.1cm),
    minimum height = \swheight,
    node distance = \baseskip,
    label={[anchor=base]above,label distance=0.7ex:{\emph{\small App}}},
  },
  mem/.style = {
    node distance = 0pt,
    draw,
    pattern=north west lines,
  },
  kmem/.style = {
    mem,
    node distance = 0pt,
    draw,
    pattern=crosshatch,
  },
  nic/.style = {
    draw,rectangle,
    node distance = 8pt,
    minimum width = (\basewidth + \basewidth + \baseskip + \baseskip),
    minimum height = 2.5ex
  },
  flow/.style = {
    rounded corners, line width = 1pt, color=red, -{Latex}
  },
  control/.style = {
    rounded corners, line width = 0.8pt, color=blue, -{Latex}
  },
  control point/.style = {
    draw,
    minimum size=6pt,
    align=center,
    inner sep=0pt,
    color=black!50,
    fill=black!50,
    node distance = 2pt
  },
  syscall send/.style = {
    control point, star, star points=3, star point ratio = 3,
  },
  send/.style = {
    control point, star, star point ratio=2 
  },
  nic send/.style = {
    control point, circle,
  },
  nic receive/.style = {
    control point, diamond
  },
  receive/.style = {
    control point
  },
  syscall receive/.style = {
    control point, regular polygon,
  },
  ghost control/.style = {
    control point, fill = none, draw = none
  }
}

\newcommand{\pict}[1]{\tikz[baseline=-2.5pt]{\node[#1](n){};}}

\definecolor{usermem}{RGB}{239,237,245}
\definecolor{nicmem}{RGB}{141,160,203}

\begin{figure}[t]
  \centering
  \includegraphics[width=\linewidth]{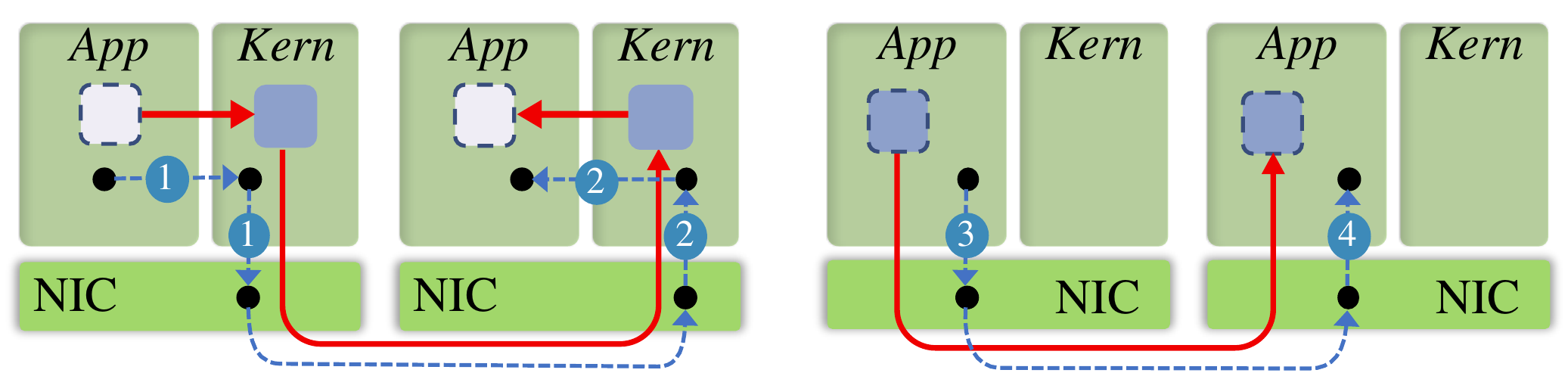}
  \renewcommand{\pict}[1]{\tikz[baseline=-2.5pt]{\node[#1,rounded corners=1pt, minimum width=0.7em,
      minimum height=0.1em, text height=0]{};}}
  \vspace{-2ex}
  \caption[Network stacks]{ Traditional (left) and RDMA (right) network stacks.
    In traditional networks, the user application triggers the NIC via
    kernel~(1). After receiving a packet, the NIC notifies the application back
    through the kernel~(2). In RDMA networks the application communicates
    directly to the NICs~(3) and vice-versa~(4) without kernel intervention.
    Traditional networks require message copy between application buffers
    (\pict{draw, densely dashed, fill=usermem}) and NIC accessible kernel buffers
    (\pict{fill=nicmem}). RDMA-NICs (right) can access the message buffer in the
    application memory directly (\pict{draw, densely dashed, fill=nicmem}).}
  \label{fig:stacks}
\end{figure}


The ability to live-migrate applications has long been available for
virtual machines~(VMs) and is widely appreciated in cloud
computing~\cite{nelsonFastTransparentMigration2005, clarkLiveMigrationVirtual2005,
  deshpandeFastServerDeprovisioning2014, hinesPostcopyLiveMigration2009,
  panCompSCLiveMigration2012}.
We expect live migration to become even more popular with the growth of
disaggregated~\cite{connorpatrickTECHNIQUESMIGRATEVIRTUAL2018,
  guEfficientMemoryDisaggregation2017},
serverless~\cite{wangReplayableExecutionOptimized2019}, and fog
computing~\cite{osanaiyeCloudFogComputing2017}.
In contrast to VMs, containerised applications share the kernel, and thus their
state, with the host system.
In general, it is still possible to extract the relevant container state from
the kernel and restore it on another host later on.
This recoverable state includes open TCP connections, shell sessions, file
locks~\cite{criu, criu-tcp}.
However, the state of RDMA communication channels is not recoverable by existing
systems, and hence applications using RDMA cannot be checkpointed or migrated.

To outline the conceptual difficulties involved in saving the state of RDMA
communication channels, we compare a traditional TCP/IP-based network stack and
the \ibverbs{} API (see~\cref{fig:stacks}).
First, with a traditional network stack, the kernel fully controls when the
communication happens: applications need to perform system calls to send or
receive a message.
In \ibverbs{}, because of direct communication between the NIC and the
application, the OS has no communication interception points, except of tearing
down the connection.
Although the OS can stop a process from sending further messages, the NIC may
still silently change the application state.
Second, part of the connection state resides at the NIC and is inaccessible for
the OS.
Creating a consistent checkpoint is impossible in this situation.

In this paper, we propose \name{}, an architecture enabling transparent live
migration of containerised RDMA-applications on the OS level.
We identify the missing hardware capabilities of existing RDMA-enabled NICs
required for transparent live migration.
We augment the underlying RoCEv2 communication protocol to update the physical
addresses of a migrated container transparently.
We modify a software RoCEv2-implementation to show that the required protocol
changes are small and do not affect the critical path of the communication.
%
%
Finally, we demonstrate an end-to-end live migration flow of containerised
RDMA-applications.



\section{Background}

This section gives a short introduction to containerisation and RDMA
networking.
We further outline live migration and how RDMA networking obstructs this
process.

\subsection{Containers}

In Linux, processes and process trees can be logically separated from the rest
of the system using \emph{namespace} isolation.
Using namespaces, allows process creation with an isolated view on the file
system, network devices, users, etc.
Container runtimes leverage namespaces and other low-level kernel
mechanisms~\cite{merkelDockerLightweightLinux2014, LinuxContainers}
to create a complete system view without external dependencies.
Considering their close relation, we use the terms container and process
interchangeably in this paper.
%
%
A distributed application may comprise multiple containers across a network: a
Spark application, for example, can run the master and each worker in an
isolated container and an MPI~\cite{openmpi} application can containerise each
\emph{rank}.


\subsection{Infiniband verbs}
\label{sec:background-verbs}

The \ibverbs{}~API is today's de-facto standard for high-performance RDMA
communication.
It enables applications to achieve high throughput and low latency by accessing
the NIC directly (\emph{OS-bypass}), avoiding unnecessary memory movement
(\emph{zero-copy}), and delegating packet processing to the NIC
(\emph{offloading}) .
%

\begin{figure}
  \centering
  \includegraphics[width=1\linewidth]{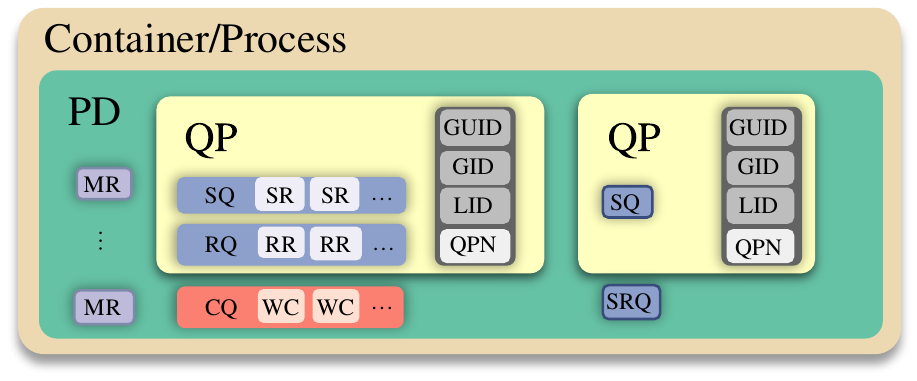}
  \caption[\ibverbs{} objects]{%
    Primitives of the \ibverbs{} library.
    Each queue pair~ (QP) comprises a send and a receive queue and has
    multiple IDs; node-global IDs (grey) are shared by all QPs on
    the same node.%
  }
  \label{fig:ibverbs}
\end{figure}


\Cref{fig:ibverbs} shows the following \ibverbs{} objects involved in
communication.
\emph{Memory regions}~(MRs) represent pinned memory shared between the
application and the NIC.
\emph{Queue pairs}~(QPs), comprising a send queue~(SQ) and a receive~(RQ) queue,
represent connections.
To reduce memory footprint, multiple QPs can replace multiple individual RQs
with use a single \emph{shared receive queue}~(SRQ).
\emph{Completion queues}~(CQs) inform the application about completed
communication requests.
A \emph{protection domain}~(PD) groups all these \ibverbs{} objects together and
represents the process address space to the NIC.

To establish a connection, an application needs to exchange the following
addressing information:
\emph{Memory protection keys} to enable access to remote MRs, the global
vendor-assigned address~(\emph{GUID}), the routable address~(\emph{GID}),
the non-routable address~(\emph{LID}), and the node-specific QP
number~(\emph{QPN}).
This exchange happens over another network, like TCP/IP.
During the connection setup, each QP is configured for a specific \emph{type of
service}.
We implement \name{} for Reliable Connections (RC) type of service, which
provides reliable in-order message delivery between two communication partners.

The application sends or receives messages by posting \emph{send
requests}~(SR) or \emph{receive requests}~(RR) to a QP.
These requests describe the message structure and refer to the memory buffers
within previously created MRs.
The application checks for the completion of outstanding work requests by
polling the CQ for \emph{work completions}~(WC).

There are various implementations of the \ibverbs{}~API for different hardware,
including Infiniband~\cite{ibta}, iWarp~\cite{iwarp}, and
RoCE~\cite{SupplementInfiniBandArchitecture2010,
  SupplementInfiniBandArchitecture2014}.
\Infiniband{} is generally the fastest among these but requires specialised NICs
and switches.
RoCE and iWarp provide RDMA capabilities in Ethernet networks.
They still require require hardware support in the NIC, however, do not depend
on specialised switches and thus make it easier to incorporate RDMA into an
existing infrastructure.
This work focuses on RoCEv2, a version of RoCE protocol.
%

To enable RDMA-application migration, it is important to consider following
challenges:
\begin{compactenum}
\item User applications have to use physical network addresses (QPN, LID, GID,
  GUID), and the \ibverbs{}~API does not specify a way for virtualising these.
\item The NIC can write to any memory it shares with the application without
the OS noticing.
\item The OS cannot instruct the NIC to pause the communication, except abruptly
  terminating it.
\item The user applications are not prepared for a connection changing
  destination address and going into an erroneous state. As, result, the
  applications will terminate abruptly.
\item Although the OS is aware of all \ibverbs{} objects created by the
  application, it does not control the whole state of these objects, as the
  state partially resides on the NIC.
\end{compactenum}
We address all of these challenges in~\cref{sec:design}.

\subsection{CRIU}
\label{sec:criu}

CRIU is a software framework for transparently checkpointing and restoring the
state of Linux processes~\cite{criu}.
It enables live migration, snapshots, or remote debugging of processes, process
trees, and containers.
To extract the user-space application state, CRIU uses conventional debugging
mechanisms~\cite{PtraceLinuxManual, ProcLinuxManual}.
However, to extract state of process-specific kernel objects, CRIU depends on
special Linux kernel interfaces.

To restore a process, CRIU creates a new process that initially runs the CRIU
executable which reads the image of the target process and recreates all OS
objects on its behalf.
This approach allows CRIU to utilise the available OS mechanisms to run most
of the recovery without the need for significant kernel modifications.
Finally, CRIU removes any traces of itself from the process.
%

CRIU is also capable of restoring the state of TCP connections.
This feature is crucial for the live migration of distributed
applications~\cite{criu-tcp}.
The Linux kernel introduced a new TCP connection state, \texttt{TCP\_REPAIR},
for that purpose.
In this state a user-level process can change the state of send
and receive message queues, get and set message sequence numbers and
timestamps, or open and close connection without notifying the other side.
%
%

As of now, if CRIU attempts to checkpoint an RDMA-application, it will detect
\ibverbs{} objects and will refuse to proceed.
Discarding \ibverbs{} objects in the naive hope that the application will be
able to recover is failure-prone: once an application runs into an erroneous
\ibverbs{} object, in most cases, the application will hang or crash.
%
Thus, we provide explicit support for \ibverbs{} objects in CRIU
(see~\cref{sec:design}).


\section{Design}
\label{sec:design}

\name{} is based on modern container runtimes and reuses much of the existing
infrastructure with minimal changes.
%
%
Most importantly, we require no modification of the software running inside the
container (see~\cref{sec:design-soft}).

Existing container runtimes rely on CRIU for checkpoint/restore
functionality~\cite{merkelDockerLightweightLinux2014, LinuxContainers, podman,
  runc}.
Therefore, it is sufficient to extend CRIU with \ibverbs{} support to checkpoint
and restore containerised RDMA-applications.
\Cref{sec:design-cr} describes our modifications to the \ibverbs{}~API and how
CRIU uses them.

We also add two new QP states to enable CRIU to create consistent checkpoints
(see~\cref{sec:design-qp}).
Finally, \cref{sec:design-cm} describes minimal changes to the packet-level
RoCEv2 protocol to ensure that each QP maintains correct information about
the location of its partner QP.
%


%

\subsection{Software Stack}
\label{sec:design-soft}

\tikzset{
  basic block/.style = {
    minimum width=1.8cm,
    inner sep = 1pt,
    node distance = 0.75cm,
  },
  block/.style = {
    draw,
    anchor=west,
    node distance = 0.5cm,
    inner sep = 1pt,
    text width=\pgfkeysvalueof{/pgf/minimum width}-2*\pgfkeysvalueof{/pgf/inner xsep},
    align = left,
  },
  modified block/.style = {
    block,
    fill = green!50,
  }
}

\newcommand{\swstacksfont}{\footnotesize \strut \hspace{0.2em}}
\newcommand{\swstacksdist}{2pt}

\begin{figure}[t]
  \centering
  \includegraphics[width=\linewidth]{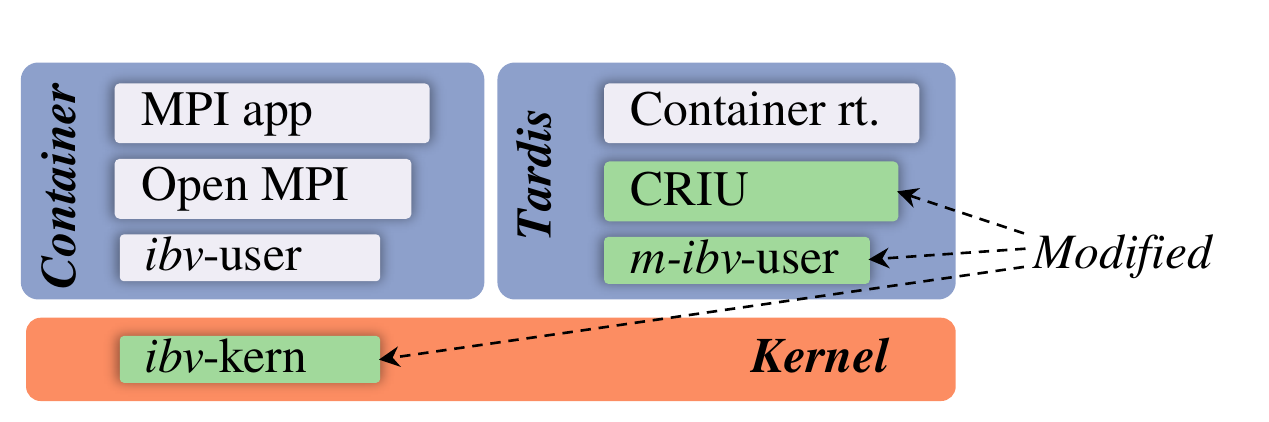}
  \caption[Parallel stacks]{Container migration architecture. Software inside
    the container, including the user level driver (\texttt{ibv}-cont, grey),
    is unmodified. The host runs CRIU, kernel (\texttt{ibv}-kern) and user
    (\texttt{ibv}-migr, green) level drivers modified for migratability.}
  \label{fig:sw-stacks}
\end{figure}


Typically, access to the RDMA network is hidden deep inside the software stack.
\Cref{fig:sw-stacks} gives an example of a containerised RDMA-application.
%
The container image comes with all library dependencies, like the libc, but not
the kernel-level drivers.
The application uses a stack of communication libraries, comprising
Open~MPI~\cite{openmpi}, Open~UCX~\cite{ucx} (not shown), and \ibverbs{}.
Normally, to migrate, container runtime would require the application inside the
container to terminate and later recover all \ibverbs{} objects.
This removes transparency from live migration.

\name{} runs alongside the container comprising of a container runtime (e.g.,
docker\cite{merkelDockerLightweightLinux2014}), CRIU, and \ibverbs{} library.
We modified CRIU to make it aware of \ibverbs{}, so that it can successfully
save \ibverbs{} objects when CRIU traverses the kernel objects belonging to the
container.
We extend the \ibverbs{} library (\emph{m-ibv}-user and \emph{ibv}-kern) to enable
serialisation and deserialisation of the \ibverbs{} objects.
Importantly, the API extension is backwards compatible with the \ibverbs{}
library running inside the container.
Thus, both \emph{m-ibv}-user and \emph{ibv}-user use the same kernel version of
\ibverbs{}.
\name{} requires no modifications of any software inside the container.

%

%
%

\subsection{Checkpoint/Restore API}
\label{sec:design-cr}

\lstdefinestyle{code}{
  language=C,
  basicstyle=\ttfamily,
}

\newcommand*{\ibvdump}{\lstinline[style=code]{ibv_dump_context}}
\newcommand*{\ibvrestore}{\lstinline[style=code]{ibv_restore_object}}

To enable checkpoint/restore for processes and containers, we extend the
\ibverbs{}~API with two new calls (see~\cref{lst:cr-api}): \ibvdump{} and
\ibvrestore{}\@.
The dump call returns dump of all \ibverbs{} objects within a specific
\ibverbs{} context.
The dumping runs almost entirely inside the kernel for two reasons.
First, some links between the objects are only visible at the kernel level.
Second, to get a consistent checkpoint it is crucial to ensure an atomic dump.

\begin{listing}
  \begin{lstlisting}[style=code, gobble=4]
    int ibv_dump_context(
            struct ibv_context *ctx,
            int *count, void *dump,
            size_t length);
    int ibv_restore_object(
            struct ibv_context *ctx,
            void **object,
            int object_type, int cmd,
            void *args, size_t length);
  \end{lstlisting}
  \caption[\ibverbs{} Checkpoint/Restart API]{Checkpoint/Restart extension for
    the \ibverbs{}~API.
    \ibvdump{} creates an image of the \ibverbs{} context
    \lstinline[style=code]{ctx} with \lstinline[style=code]{count}
    objects and stores it in the caller-provided memory region
    \lstinline[style=code]{dump} of size \lstinline[style=code]{length}.
    \ibvrestore{} executes the restore command \lstinline[style=code]{cmd} for
    an individual object (QP, CQ, etc.) of type
    \lstinline[style=code]{object_type}.
    The call expects a list of arguments specific to the object type and
    recovery command.
    \lstinline[style=code]{args} is an opaque pointer to the argument buffer of size
    \lstinline[style=code]{length}.
    A pointer to the restored object is returned via
    \lstinline[style=code]{object}. }
  \label{lst:cr-api}
\end{listing}

Of course, the existing \ibverbs{} API allows to create new objects.
However, the existing \ibverbs{}~API is not expressive enough for
\emph{restoring} objects.
For example, when restoring a completion queue~(CQ), the current API does not
allow to specify the address of the shared memory region for this queue.
Also it is not possible to recreate a queue pair~(QP) directly in the
Ready-to-Send~(RTS) state.
Instead, the QP has to traverse all intermediate states before reaching RTS.
 

We introduce a fine-grained \ibvrestore{} call to restore \ibverbs{} objects one
by one, for situations when the existing API is not sufficient.
In turn, modified \name{}, uses the extended \ibverbs{} API to save and restore the
\ibverbs{} state of applications.
During recovery, \name{} reads the object dump and applies a specific recovery
procedure for each object type.
For example, to recover a QP, \name{} calls \ibvrestore{} with the command
\lstinline[style=code]{CREATE} and progresses the QP through the Init, RTR,
and RTS states using \lstinline[style=code]{ibv_modify_qp}\@.
The contents of memory regions or QP buffers are recovered using the standard
file and memory operations.
Finally, \name{} brings the queue to the original state using the
\lstinline[style=code]{REFILL} command of the restore call.


\subsection{Queue Pair States}
\label{sec:design-qp}
\definecolor{new state border}{RGB}{247,205,49}
\definecolor{new state fill}{RGB}{255,255,191}
\definecolor{error state border}{RGB}{252,141,89}
\definecolor{error state fill}{RGB}{253,187,157}
\definecolor{state border}{RGB}{107,166,202}
\definecolor{state fill}{RGB}{145,191,219}

\tikzset{
  state shadow/.style = {
    drop shadow = {opacity = 0.3}
  },
  basic/.style = {
    inner sep = 0pt,
    line width=0.03cm,
    font ={\footnotesize{#1}},
  },
  state attribute/.style={
    draw,
    color=state border,
    fill=state fill,
    text=black,
    circle,
    state shadow,
  },
  state/.style={
    state attribute,
    node distance= 2.2cm,
    minimum size=2em,
    line width=0.03cm,
    basic,
  },
  error attribute/.style = {
    line width=0.03cm,
    densely dotted,
    color = error state border,
    fill = error state fill,
    text = black,
  },
  new attribute/.style = {
    color = new state border,
    fill = new state fill,
    text = black,
    line width=0.03cm,
    densely dashed
  },
  error state/.style = {
    state,
    thick,
  	error attribute,
  },
  new state/.style = {
    state,
    new attribute,
  },
  transition/.style = {
    draw,
    color=state border,
    -{Latex},
    >=Latex,
  },
  error transition/.style = {
    transition,
  	error attribute,
    fill=none,
  },
  new transition/.style = {
    transition,
    new attribute,
    fill=none,
  },
}

\newcommand{\arrowpict}[1]{\tikz[baseline=-2.5pt,x=1em,y=1em]{\draw[#1, ->](0,0) -- (1,0);}}

\begin{figure}
  \centering
  \begin{tikzpicture}
  \node (ne) {};
  \node[basic] at (ne) [yshift=0.2cm] {Create};
  \node[state] (r) [node distance=3em, right of = ne] {R};
  \node[state] (init) [node distance=5em, below of =  r] {Init};
  \node[state] (rtr) [right of = init] {RTR};
  \node[state] (rts) [right of = rtr] {RTS};
  \node[error state] (e) [right of = r] {E};
  \node[error state] (sqe) [right of = e] {SQE};
  \node[state] (sqd) [xshift=0.9cm, yshift=0.8cm] at (rtr) {SQD};
  \node[new state] (s) [right of = sqe] {S};
  \node[new state] (p) [right of = rts] {P};

  \draw[transition] (ne) -> (r);
  \draw[transition] (e) -> (r);
  \draw[transition] (r) -> (init);
  \draw[transition] (init) -> (rtr);
  \draw[transition] (rtr) -> (rts);
  \draw[transition]  (rts) edge[<->] (sqd);
  \draw[transition] (sqe) edge[bend right=0] (rts);

  \draw (init) edge [transition, loop left] (init);
  \draw (rts) edge [transition, loop right] (rts);
  \draw (sqd) edge [transition, loop left] (sqd);

  \draw[error transition] (init) -> (e);
  \draw[error transition] (rtr) -> (e);
  \draw[error transition] (sqd) -> (e);
  \draw[error transition] (sqd) -> (sqe);
  \draw[error transition] (rts) to[bend right=20]  (e);
  \draw[error transition] (sqe) -> (e);
  \draw[error transition] (rts) to[bend right] (sqe);

  \draw[new transition] (rts) -> (s);
  \draw[new transition] (rts) to[bend right=25] (p);
  \draw[new transition] (p) to[bend right=25] (rts);
\end{tikzpicture} 
  \caption[QP State Diagram]{QP State Diagram.
    Normal states and state transitions (\pict{state attribute},
    \arrowpict{transition}) are controlled by the user application.
    A QP is put into error states (\pict{draw, circle, error attribute, state shadow},
    \arrowpict{error transition}) either by the OS or the NIC.
    New states (\pict{draw, circle, new attribute, state shadow},
    \arrowpict{new transition}) are used for connection migration.
   }
  \label{fig:qp-states}
\end{figure}
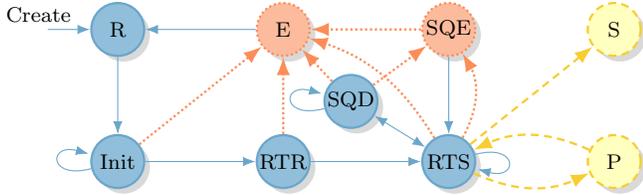


Before communication can commence, an application establishes a connection
bringing a QP through a sequence of states (depicted in~\cref{fig:qp-states}).
Each newly-created QP is in the \emph{Reset}~(R) state.
To send and receive messages, a QP must reach its final
\emph{Ready-to-Send}~(RTS) state.
Before reaching RTS, the QP traverses the \emph{Init} and
\emph{Ready-to-Receive}~(RTR) states.
In case of an error, the QP goes into one of the error states; \emph{Error}~(E),
or \emph{Send Queue Error}~(SQE).
In the \emph{Send Queue Drain}~(SQD) state, a QP does not accept new send
requests.
Apart from that, SQD is equivalent to the RTS state.

In addition to the existing states, we add two new states invisible to the user
application (see~\cref{fig:qp-states}): \Sstopped{}~(S) and \Spaused{}~(P).
When the kernel executes \ibvdump{}, all QPs of the specified context go into a
\Sstopped{} state.
A stopped QP does not send or receive any messages.
The QPs remain stopped until they are destroyed together with the checkpointed
process.

A QP becomes \Spaused{} when learns its destination QP has become \Sstopped{}
(see~\Cref{sec:design-cm}).
A paused QP does not send messages, but also has no other QP to receive messages
from.
A QP remains paused, until the migrated destination QP restores at a new
location and sends a message with the new location address.
The paused QP retains the new location of the destination QP and returns to
RTS state.
After that, the communication can continue.
 

\subsection{Connection Migration}
\label{sec:design-cm}

\definecolor{ready fill}{RGB}{145,191,219}
\definecolor{stopped border}{RGB}{247,205,49}
\definecolor{stopped fill}{RGB}{255,255,191}
\definecolor{invalid fill}{RGB}{253,187,157}

\tikzset{%
   state/.style={draw, circle, inner sep=1pt, minimum size=1em, font=\footnotesize},
    rts/.style={state attribute, state},
    stopped/.style={new attribute, state, state shadow},
    invalid/.style={error attribute, state, state shadow}
}

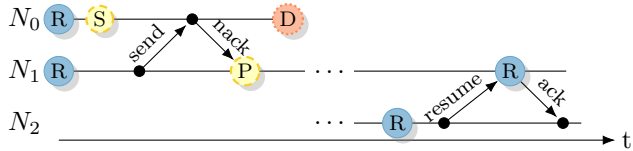
\begin{figure}[t]
  \centering
  \begin{tikzpicture}[node distance=0.15cm]
    \node (s) {$N_0$};
    \node (p) [below = of s] {$N_1$};
    \node (d) [below = of p] {$N_2$};

    \coordinate (s-start) at ($(s)+(3ex,0)$);
    \coordinate (p-start) at ($(p)+(3ex,0)$);
    \coordinate (d-start) at ($(d)+(3ex,0)$);

    \draw (s-start) -- ($(s)+(3.5cm,0)$) ;
    \draw (p-start) -- ($(p)+(7.2cm,0)$) node[xshift=1ex] {};
    \draw ($(d-start)+(3.5cm,0)$) -- ($(d)+(7.4cm,0)$) node[xshift=1ex] {};
    \draw[-{Latex}] ($(d-start)+(0,-1.5ex)$) -- ($(d-start)+(7.4cm,-1.5ex)$) node[xshift=1ex] {t};

    \node[rts] (s-rts) at (s-start) {R};
    \node[stopped] (stopped) [right = of s-rts] {S};

    \coordinate (s-rcv) at ($(stopped) + (3.5em,0)$);
    \fill[fill=black,draw] (s-rcv) circle (2pt);

    \node[invalid] (invalid) at ($(s) + (3.5cm,0)$)  {D};

    \node[rts] (p-rts) at (p-start) {R};
    \coordinate (p-send) at ($(p-start-|s-rcv) - (2em,0)$) {};
    \fill[fill=black,draw] (p-send) circle (2pt);
    \node[stopped] (wait) at ($(p-send) + (4em,0)$) {P};
    \coordinate (pre resume) at ($(p-start)+(6cm, 0)$);
    \node[rts] (resume) at (pre resume) {R};

    
    \node[rts] (restore) at ($(d-start) + (4.5cm,0)$) {R};
    \node[fill=white] (pre-init) at ($(restore) - (0.9cm,0)$) {$\ldots$};
    \coordinate (send-resume) at ($(resume|-restore) - (2.5em,0)$);
    \fill[fill=black,draw] (send-resume) circle (2pt);
    \coordinate (resume-ack) at ($(send-resume) + (4.5em,0)$);
    \fill[fill=black,draw] (resume-ack) circle (2pt);

    \node[fill=white] (pre-init) at (pre-init|-p) {$\ldots$};


    \path[draw,-{Latex},shorten >=3pt] (p-send) -> (s-rcv) node[sloped,above, near start,xshift=0.5ex] {\footnotesize send};
    \path[draw,-{Latex}] (s-rcv) -> (wait) node[midway,sloped,above,near end,xshift=-0.3ex] {\footnotesize nack};
    \path[draw,-{Latex}] (send-resume) -> (resume) node[midway,sloped,above,near start,xshift=0.5ex] {\footnotesize resume};
    \path[draw,-{Latex},shorten >=3pt] (resume) -> (resume-ack) node[midway,sloped,above] {\footnotesize ack};
  \end{tikzpicture}
  \renewcommand{\pict}[2]{\tikz[baseline=-2.5pt]{\node[#1] {#2};}}
  \caption[Pause/resume protocol]{To migrate from host~$N_0$ to host~$N_2$,
    the state of the QP changes from RTS~(\pict{rts}{R}) to Stopped~%
    (\pict{stopped}{S}).
    Finally, the QP is destroyed~(\pict{invalid}{D}).
    If the partner QP at host~$N_1$ sends a message during
    migration, this QP gets paused~(\pict{stopped}{P}).
    Both QPs resume normal operation once the migration is complete.
  }
  \label{fig:pause-resume}
\end{figure}


There are two considerations, when migrating a connection.
First, during the migration, the communication partner of the migrating
container must not confuse migration with a network failure.
Second, once the migration is complete, all partners of the communication node
need to learn its new address.
%

We address the first issue by extending RoCEv2 with a connection migration
protocol.
The connection migration protocol is active during and after migration
(see~\cref{fig:pause-resume}).
This protocol is part of the low-level packet transmission protocol and is
typically implemented entirely within the NIC.
Also, we add a new negative acknowledgement type \nakstopped{}.
If a stopped QP receives a packet, it replies with \nakstopped{} and drops the
packet.
%
%
When the partner QP receives this negative acknowledgement, it transitions to
the \Spaused{}~(P) state and refrains from sending further packets until
receiving a resume message.

After migration completes, the new host of the migrated process restores all QPs
to their original state.
Once a QP reaches the RTS state, the new host executes the
\lstinline[style=code]{REFILL} command.
This command restores the driver-specific internal QP state and sends a newly
introduced \emph{resume} message to the partner QP.
Resume messages are sent unconditionally, even if the partner QP was not paused
before.
%
This way, we also address the second issue:
The recipient of the resume message updates its internal address information to
point to the new location of the migrated QP; the source address of the resume
message.
%

Each pause and resume message carry source and destination information.
Thus, if multiple QPs migrate at the same time, there can be no confusion which
QPs must be paused or resumed.
If at any point the migration process fails, the paused QPs will remain stuck
and will not resume communication.
This scenario is completely analogous to a failure during a TCP-connection
migration.
In both cases, \name{} will be responsible for cleaning up the resources.



%
%
%
%


\section{Implementation}

To provide transparent live migration, \name{} incorporates changes
to CRIU, \ibverbs{} library, RDMA-device driver (SoftRoCE), and packet-level
RoCEv2-protocol.
To migrate an application, the container runtime invokes CRIU which checkpoints
the target container.
CRIU stops active RDMA-connections and saves the state of \ibverbs{} objects
(see~\cref{sec:criu-impl}).
SoftRoCE then pauses communication using our extensions to the packet-level
protocol.
After transfering the checkpoint to the destination node, the container runtime
at that node invokes CRIU to recover the \ibverbs{} objects and restores the
application.
SoftRoCE then resumes all paused communication to complete the migration
process.

SoftRoCE is a Linux kernel-level software implementation (not an
emulation~\cite{liranlissLinuxSoftRoCEDriver2017}) of the RoCEv2
protocol~\cite{SupplementInfiniBandArchitecture2014}.
RoCEv2 runs RDMA communication by tunnelling Infiniband packets through a
well-known UDP port.
%
%
%
In contrast to other RDMA-device drivers, SoftRoCE allows the OS to inspect,
modify, and control the state of \ibverbs{} objects completely.
%

As a performance-critical component of RDMA communication, RoCEv2 usually runs
in NIC hardware.
So changes to the protocol require hardware changes.
We implement \name{} with the focus on minimising these protocol changes.
The key part of \name{} is the addition of connection migration capabilities to
the existing RoCEv2 protocol (see~\cref{sec:softroce}).
%
%
%

\subsection{State Extraction and Recovery}
\label{sec:criu-impl}

State extraction begins when CRIU discovers that its target process opened an
\ibverbs{} device.
We modified CRIU to use the API presented in~\cref{sec:design-cr} to extract the
state of all available \ibverbs{} objects.
CRIU stores this state together with other process data in an image.
Later, CRIU recovers the image on another node using the new API.

When CRIU recovers MRs and QPs of the migrated application, the recovered
objects must maintain their original unique identifiers.
These identifiers are system-global and assigned by the NIC (in our case the
SoftRoCE driver) in a sequential manner.
We augmented the SoftRoCE driver to expose the IDs of the last assigned MR and
QP to \name{} in userspace.
These IDs are \emph{memory region number} (MRN) and \emph{queue pair number}
correspondingly.
Before recreating an MR or QP, CRIU configures the last ID appropriately.
If no other MR or QP occupies this ID, the newly created object will maintain
the original ID.
This approach is analogous to the way CRIU maintains the process ID of a
restored process using \lstinline[language=bash]{ns_last_pid} mechanism in
Linux, which exposes the last process ID assigned by the kernel.

It is possible for some other process to occupy MRN or QPN, which CRIU wants to
restore.
Two processes cannot use the same MRN or QPN on the same node, resulting in a
conflict.
In the current scheme, we avoid these conflicts by partitioning QP and MR
addresses globally among all nodes in the system before the application startup.
CRIU faces the very same problem with process ID collisions.
This problem has only been solved with the introduction of process ID
namespaces.
To remedy the collision problem for \ibverbs{} objects, a similar
namespace-based mechanism, would be required.
We leave this issue for future work.

Additionally, recovered MRs, have to maintain their original memory protection
keys.
The protection keys are pseudo-random numbers provided by the NIC and are used
by a remote communication partner when sending a packet.
An RDMA operation succeeds only if the provided key matches the expected key of
a given MR.
Other than that, the key's value does not carry any additional semantics.
Thus, no collision problems exist for protection keys.

CRIU sets all protection keys to their original values before communication
restarts by making an \lstinline[language=C]{ibv_restore_object} call with the
\lstinline[language=C]{IBV_RESTORE_MR_KEYS} command.
%
%

\subsection{Resuming Connections}
\label{sec:softroce}

The connection migration protocol ensures that connections are terminated
gracefully and recovered to a consistent state.
The implementation of this protocol is device- and driver-specific.
In this work, we modify the SoftRoCE driver to make it compliant with the
connection migration protocol (\cref{sec:design-cm}) by providing an
implementation of the checkpoint/restore~API (\cref{sec:design-cr}).

\definecolor{comppkt}{RGB}{161,215,106}
\definecolor{ackpkt}{RGB}{255,255,191}
\definecolor{unackpkt}{RGB}{253,187,157}

\tikzset{%
	steplbl/.style = {circle, fill, inner sep=.42pt, outer sep=0pt, text=white, node font=\footnotesize, minimum size=2.3ex},
	ssteplbl/.style = {steplbl, star, star points=7, draw, fill=white, ultra thick},
  buf/.style = {draw, minimum height=.7\baseHeight, minimum width=.2\baseWidth},
  queue item/.style = {draw, node distance=1pt, text width=4em, text height=2ex, node font=\footnotesize, inner sep=2pt},
  lbl/.style = {outer sep=0pt, inner sep=0pt},
  frame/.style = {draw},
  wr/.style = {draw, node distance=1pt, minimum width=2.1em, text height=1.5ex, minimum height=1.7em, inner xsep=0, outer xsep=0, node font=\footnotesize},
  pkt/.style = {draw, node distance=0, minimum width=1.5em, text height=1.3ex, inner xsep=0, outer xsep=0, node font=\footnotesize},
  leg-shadow/.style = {rounded corners=0.5mm,
    blur shadow ={shadow xshift = 0, shadow yshift = 0.0em, shadow scale = 1.2}},
  pkt-done/.style = {pkt, color=white, fill = comppkt,   leg-shadow},
  pkt-sent/.style = {pkt, color=white, fill = ackpkt,    leg-shadow},
  pkt-pending/.style = {pkt, color=white, fill=unackpkt, leg-shadow},
}

\begin{figure}
  \centering
  \includegraphics[width=\linewidth]{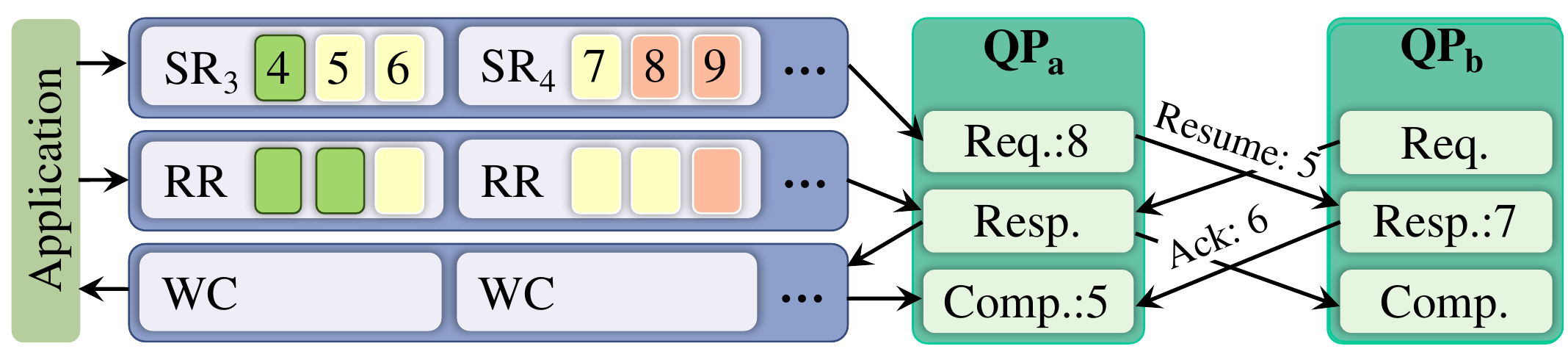}
  \renewcommand{\pict}[1]{\tikz[baseline=-2.5pt]{\node[#1,minimum width=0.7em,
      minimum height=0.1em, text height=0]{};}}
  \caption[Connection resume]{Resuming connection in SoftRoCE. Packets 8,
    9~(\pict{pkt-pending}) are to be processed by the requester. Packets
    5-7~(\pict{pkt-sent}) are yet to be acknowledged. Packet 4~(\pict{pkt-done})
    is already acknowledged. $\text{QP}_b$ expects the next packet is 7. Resume
    packet has PSN of the first unacknowledged packet~(\pict{pkt-sent}).
    $\text{QP}_b$ replies with an acknowledgement of the last received packet.}
  \label{fig:softroce-resume}
\end{figure}


\Cref{fig:softroce-resume} outlines the basic operation of the SoftRoCE driver.
The driver creates three kernel tasks for each QP: \emph{requester},
\emph{responder}, and \emph{completer}.
When an application posts send~(SR) and receive~(RR) work requests to a QP, they
are processed by requester and responder correspondingly.
A work request may be split into multiple packets, depending on the MTU size.
When the whole work request is complete, responder or completer notify the
application by posting a work completion to the completion queue.

The kernel tasks process all requests packet by packet.
Each task maintains the packet sequence number~(PSN) of the next packet.
A packet sent by a requester is processed by the responder of the partner QP.
The responder replies with an acknowledgement that is processed by the
completer.
The completer generates a work completion~(WC) after receiving acknowledgement
for the last packet in an SR.
Similarly, the responder generates a WC after receiving all packets of an RR.

{
\newcommand*{\QP}[1]{QP\textsubscript{#1}}

After migration, when the recovered \QP{a} is ready to communicate again, it
sends a resume message to \QP{b} with the new address.
This way, \QP{b} learns the new location of \QP{a}.
Receiving this resume message, the responder of \QP{b} replies with
an acknowledgement of the last successfully received packet.
If some packets were lost during the migration, the next PSN at the responder of
\QP{b} is smaller than the next PSN at the requester of \QP{a}.
The difference corresponds to the lost packets, which must be retransmitted.
Simultaneously, the requester of \QP{b} can already start sending messages.
At this point, the connection between \QP{a} and \QP{b} is fully
recovered.
}

The presented protocol ensures that both QPs recover the connection without
losing packets irrecoverably.
If packets were lost during migration, the QPs can determine which packets were
lost and retransmit them.
This retransmission is part of the normal RoCEv2 protocol.
The whole connection migration protocol runs transparently for the user
applications.


\section{Evaluation}

We evaluate \name{} from three main aspects.
First, we analyse the implementation effort, with a specific focus on changes to
the RoCEv2 protocol.
%
%
Second, we study the overhead of adding migration capability, outside of the
migration phase.
Third, we estimate the fine-grained cost of migration for individual \ibverbs{}
objects, as well as the full latency of migration in realistic
RDMA-applications.

For most experiments, we use a system with two machines:
Each machine is equipped with an Intel i7-4790~CPU, \SI{16}{\gibi\byte}~RAM, an
on-board Intel \SI{1}{Gb}~Ethernet adapter, a Mellanox ConnectX-3 VPI
adapter, and a Mellanox Connect-IB \SI{56}{Gb} adapter.
The Mellanox~VPI adapters are set to \SI{40}{Gb} Ethernet mode and
connected to a Cisco C93128TX \SI{40}{Gb} Ethernet switch.
The SoftRoCE driver communicates over this adapter.
The machines run Debian~11 with a custom Linux~5.7-based kernel.
We refer to this setup as \emph{local}.

We conduct further measurements on a cluster comprising two-socket Intel
E5-2680~v3 CPUs nodes with Connect-IB \SI{56}{Gb} NICs deployed by Bull.
We refer to this setup as \emph{cluster}.
Two nodes similar to those in the cluster were used in a local setup and
equipped with Mellanox ConnectX-3 VPI NICs configured to \SI{56}{Gb}
InfiniBand mode.

\subsection{Magnitude of Changes}

\begin{table}
  \centering
  \begin{tabular}{lln{6}{0}n{4}{0}}
    \toprule
    Level & Component          & Original & {{$\Delta$}} \\
    \midrule
    Kernel & \ibverbs{}& 30565 & 719  \\
              & SoftRoCE  & 9446 & 872  \\
              & QP tasks & 1112 & 249 \\ 
    User & \ibverbs{} & 12431 & 339 \\
              & SoftRoCE   & 1004  & 332 \\
              & CRIU        & 61616 & 1845 \\
    \cmidrule{4-4}
    Total      &            &       & 4137 \\
    \bottomrule
  \end{tabular}
  \caption{Development effort in SLOC. We specifically show magnitude of changes
    done to QP tasks (see~\cref{fig:softroce-resume}).}
  \label{tab:sloc}
\end{table}

\name{} requires few changes to the low-level RoCEv2 protocol, as shown in
\cref{tab:sloc}.
We count newly added or modified source lines of code in different components of
the software stack.
Only around 10\% of all the changes apply to the kernel-level SoftRoCE driver.
These changes mostly focus on saving and restoring the state of \ibverbs{}
objects.
We counted separately changes to the requester, responder, and completer QP
tasks responsible for the active phase of communication
(see~\cref{fig:softroce-resume}).
Such QP tasks are often implemented in the NIC hardware, for other
RDMA-implementations.
Therefore it is important to minimise changes specifically to QP tasks, as
changes there directly translate to hardware changes.
In our implementation, changes to QP tasks accounted only for around 6\% of
overall changes.

We used \texttt{gprof} to record the coverage of connection migration support
code outside of migration phase.
Out of all changes done to the QP tasks, only 28 lines were touched, while the
application communication was active.
Among them, 3 lines are variable assignments, one is an unconditional jump, the
rest are newly introduced if-else-conditions that occur at most once per packet
sent or received.
The rest of the code changes to the QP task run only during the connection
migration phase.

\begin{table}
  \centering
  \begin{tabular}{llr}
    \toprule
    Object    & Features required & State (b) \\
    \midrule
    PD        & None              & 12   \\
    MR        & Set MR keys and MRN   & 48   \\
    CQ        & Set ring buffer state  & 64   \\
    SRQ       & Set ring buffer state & 68   \\
    QP        & + QP tasks state & 271  \\
              & \qquad{} , set QPN & \\
    QP w/ SRQ & + Current WQE state  & 823  \\
    \bottomrule
  \end{tabular}
  \caption{Additional features implemented in the kernel-level SoftRoCE driver
    to enable recovery of \ibverbs{} objects. We provide the size each object
    occupies in the dump.}
  \label{tab:object-sizes}
\end{table}

Besides additional logic to the QP tasks, saving and restoring \ibverbs{}
objects requires manipulation of implementation-specific attributes.
Some of these attributes cannot be set through original \ibverbs{} API.
For example, recovery of an MR requires an additional ability to restore the
original values of memory keys and an MRN.
Some other attributes are not visible in original \ibverbs{} API at all.
The queues (CQ, SRQ, QP) implemented in SoftRoCE require an ability to save and
restore metadata of ring buffers backing up the queues.
If a QP uses a shared receive queue (SRQ), the dump of the QP additionally
includes the full state of the current \emph{work queue entry} (WQE).
We identified all required attributes for SoftRoCE, calculated their memory
footprint~(see~\Cref{tab:object-sizes}), and implemented features required by
these attributes.


We show the analysis of the required changes to RoCEv2 implemented by SoftRoCE.
We claim that similar changes are required to other low-level implementations of
RoCEv2 protocol residing in RDMA-capable NICs.
We demonstrate the changes to the communication path are minimal, outside of the
migration phase.
We reasonably expect that once mapped to the hardware the proposed changes will
remain minimal.

\subsection{Overhead of Migratability}
\label{sec:eval-migratability}

Just adding the capability for transparent container migration already may incur
overhead even when the migration does not occur.
For example, DMTCP (see~\Cref{sec:related}) intercepts all \ibverbs{} library
calls and rewrites both work requests and completions before forwarding them to
the NIC.
The interception happens persistently, even when the process running under DMTCP
never migrates.
In contrast to this, \name{} does not intercept communication operations at the
critical path, thereby introducing no measurable overhead.
This subsection explores the overhead added for normal communication operations
without migrations.

\begin{figure}
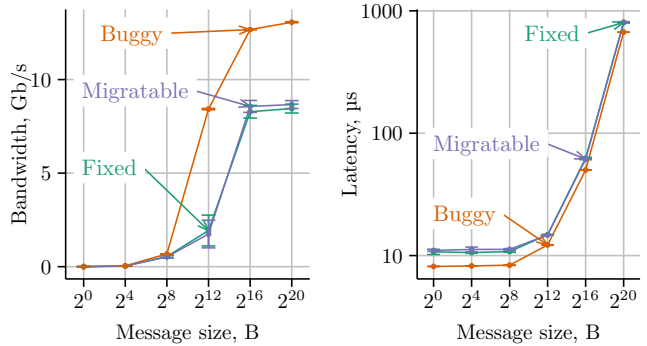

  \begin{subfigure}[t]{0.5\linewidth}
    \centering \input{rxe.bw.tex}
    \centering
    \caption{Communication Throughput}
    \label{fig:rxe-bw}
  \end{subfigure}%
  ~ 
  \begin{subfigure}[t]{0.5\linewidth}
    \centering \input{rxe.lat.tex}
    \centering
    \caption{Communication Latency}
    \label{fig:rxe-lat}
  \end{subfigure}%
  \caption[SoftRoCE Performance]{Performance comparison of different SoftRoCE
    drivers.
    The original version shows a better performance whereas adding
    connection migration support to the modified version makes practically
    no impact.
    %
  }
  \label{fig:rxe}
\end{figure}


First, we reaffirm that the proposed low-level protocol changes are minimal.
For that, we need to compare performance of migratable and non-migratable
versions of SoftRoCE driver.
Unfortunately, the original version (\emph{vanilla kernel}, without any
modifications from our side) of the SoftRoCE driver turned out to be notoriously
unstable.\hspace{-.08em}\footnote{\texttt{SIGINT} to a user-level RDMA-application caused
  the kernel to panic.}
The original driver contained multitude of concurrency bugs and required
significant restructuring.

Finally, we ended up with three versions of the driver: the original
\emph{buggy} version, a non-migratable fixed version, and a migratable fixed
version (see~\cref{fig:rxe}).
The original version rendered to be faster, nevertheless for the scope of our
paper correctness was of higher priority than the performance.
Nevertheless, the performance of both fixed versions of the SoftRoCE driver is
practically indistinguishable.
Therefore, we conclude that \name{} introduces no runtime overhead outside of
the migration phase.

\begin{figure}
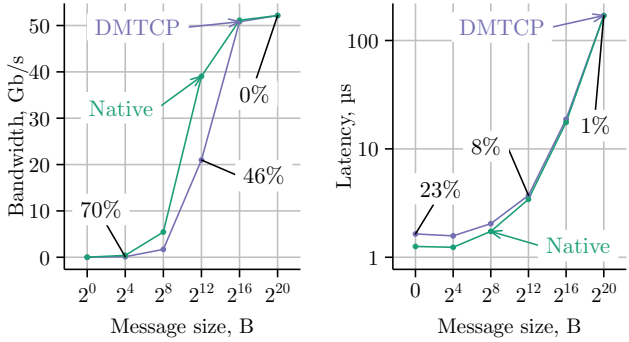

  \begin{subfigure}[t]{0.47\linewidth}
    \centering \input{dmtcp.bw.tex}
    \centering
    \caption{Communication throughput. DMTCP reduces the bandwidth by up to
      70\% for small messages.}
    \label{fig:dmtcp-bw}
  \end{subfigure}%
  ~~ 
  \begin{subfigure}[t]{0.47\linewidth}
    \centering \input{dmtcp.lat.tex}
    \centering
    \caption{Communication latency. DMTCP increases the latency by up to 23\% or
      \SI{0.34}{\us}, if $\text{size} < \SI{16}{\kibi\byte}$, else
      \SI{1.3}{\us}.}
    \label{fig:dmtcp-lat}
  \end{subfigure}%
  \caption[DMTCP Microbenchmarks]{DMTCP adds substantial communication overhead,
    even when migration is not used.
  }
  \label{fig:dmtcp}
\end{figure}


Next, we show the overhead added by DMTCP, which intercepts all \ibverbs{}
calls.
This way, we study the cost of adding migration capability at the user level.
We use the latency and bandwidth benchmarks from the OSU~5.6.1 benchmark
suite~\cite{MVAPICHBenchmarks} running on top of Open~MPI~4.0~\cite{openmpi}.
We ran the experiment on the previously described \emph{cluster} with ConnectIB
NICs.
As we have shown above, adding support for the migration does not add
performance penalty.
Thus, running without DMTCP is similar to having native migration support.
To be able to extract the state of \ibverbs{} objects, DMTCP maintains
\emph{shadow objects}, which act as proxies between the user process and the
NIC~\cite{caoTransparentCheckpointrestartInfiniband2014}.
\Cref{fig:dmtcp} shows that maintaining these shadow objects incurs a
non-negligible runtime overhead for RDMA networks.
%


%


\subsection{Migration Costs}
\label{sec:eval-costs}

\begin{table}
  \centering
  \begin{tabular}{llr}
    \toprule
    Short  & Full name                                & Location \\
    \midrule{}
    SR     & SoftRoCE                                 & local \\
    CX3/40 & ConnectX-3 \SI{40}{Gb} Ethernet   & local  \\
    CX3/56 & ConnectX-3 \SI{56}{Gb} InfiniBand & cluster \\
    CIB    & ConnectIB                                & local \\
    BIB    & Bull Connect-IB                          & cluster \\
    \bottomrule
  \end{tabular}
  \caption{RDMA-capable NICs used for the evaluation.}
  \label{tab:nics}
\end{table}

With added support for migrating \ibverbs{} objects, the container migration
time will increase proportionally to the time required to recreate these
objects.
Our goal is to estimate the additional latency for migrating RDMA-enabled
applications.
This subsection shows the cost for migrating connections created by SoftRoCE, as
well as the cost for connection creation with hardware-based \ibverbs{}
implementations.

\begin{figure}
  \centering
  \input{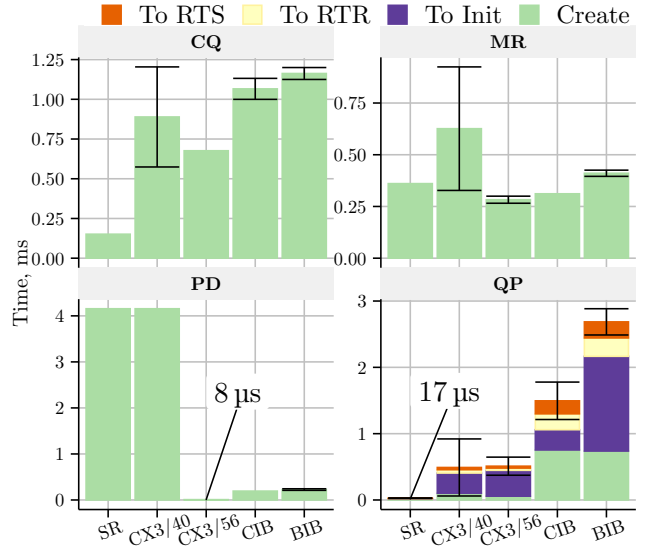}
  \caption[Object Creation]{Object creation time for different RDMA-devices.
    Before being able to send a message, a QP needs to be in the state RTS,
    which requires the traversal of three intermediate states (Reset, Init,
    RTR). We show the interval of the standard deviation around the mean.}
  \label{fig:obj}
\end{figure}


Several \ibverbs{} objects are required before a reliable connection (RC) can
be established, see \cref{sec:background-verbs}.
Usually, an application creates a single PD, one or two CQs, multiple memory
regions, and one QP per communication partner.

To measure the cost of creating individual \ibverbs{} objects, we modified
\sendbw{} from the \perftest{} benchmark suite to create additional MR objects.
We created one CQ, one PD, 64 QPs, and 64 \SI{1}{\mebi\byte}-sized MRs per run.
\Cref{fig:obj} shows the average time required to create each object across
50~runs.
%
%
Each tested NIC is represented by a bar.

We draw two conclusions from this experiment.
First, there is a substantial variation for all operations across different
NICs.
Second, the time required for most operations is in the range of milliseconds.

The exact time required for migrating RDMA connections depends on two factors:
the number of QPs and the total amount of memory assigned to
MRs~\cite{mietkeAnalysisMemoryRegistration2006}.
Both of these factors are application-specific and can vary greatly.
Therefore, next we show how the migration time is influenced by the
application's usage of MRs and QPs.

\begin{figure*}[t]
  \minipage{0.32\textwidth}
  \centering
  \input{mrs-eval.tex}
  \caption[Creating MRs]{MR registration time\\depending on the region size.}
  \label{fig:mrs}
  \endminipage~
  \minipage{0.27\textwidth}
  \centering
  \input{qps.tex}
  \caption[Migrating QPs]{Migration speed\\ with different numbers of QPs.}
  \label{fig:qps}
  \endminipage\hfill
  \minipage{0.38\textwidth}
  \centering
  \input{docker.tex}
  \vspace{-4mm}
  \caption{Migration speed comparison of Docker against CR-X (X)}
  \label{fig:docker}
  \endminipage\hfill
  \begin{tikzpicture}[remember picture, overlay,x=1pt,y=1pt,yshift=-1.6cm,xshift=-16.5cm]
    \begin{scope}
      \definecolor{fillColor}{RGB}{255,255,255}

      \path[fill=fillColor] ( 31.85,118.70) rectangle (228.95,127.24);
    \end{scope}
    \begin{scope}
      \definecolor{fillColor}{RGB}{255,255,255}

      \path[fill=fillColor] ( 37.85,118.70) rectangle ( 46.38,127.24);
    \end{scope}
    \begin{scope}
      \definecolor{fillColor}{RGB}{27,158,119}

      \path[fill=fillColor] ( 42.11,122.97) circle (  1.96);
    \end{scope}
    \begin{scope}
      \definecolor{drawColor}{RGB}{27,158,119}

      \path[draw=drawColor,line width= 0.6pt,line join=round] ( 38.70,122.97) -- ( 45.53,122.97);
    \end{scope}
    \begin{scope}
      \definecolor{fillColor}{RGB}{255,255,255}

      \path[fill=fillColor] ( 72.60,118.70) rectangle ( 81.14,127.24);
    \end{scope}
    \begin{scope}
      \definecolor{fillColor}{RGB}{217,95,2}

      \path[fill=fillColor] ( 76.87,126.02) --
      ( 79.51,121.45) --
      ( 74.23,121.45) --
      cycle;
    \end{scope}
    \begin{scope}
      \definecolor{drawColor}{RGB}{217,95,2}

      \path[draw=drawColor,line width= 0.6pt,line join=round] ( 73.45,122.97) -- ( 80.28,122.97);
    \end{scope}
    \begin{scope}
      \definecolor{fillColor}{RGB}{255,255,255}

      \path[fill=fillColor] (107.47,118.70) rectangle (116.00,127.24);
    \end{scope}
    \begin{scope}
      \definecolor{fillColor}{RGB}{117,112,179}

      \path[fill=fillColor] (109.77,121.01) --
      (113.70,121.01) --
      (113.70,124.93) --
      (109.77,124.93) --
      cycle;
    \end{scope}
    \begin{scope}
      \definecolor{drawColor}{RGB}{117,112,179}

      \path[draw=drawColor,line width= 0.6pt,line join=round] (108.32,122.97) -- (115.15,122.97);
    \end{scope}
    \begin{scope}
      \definecolor{fillColor}{RGB}{255,255,255}

      \path[fill=fillColor] (155.77,118.70) rectangle (164.31,127.24);
    \end{scope}
    \begin{scope}
      \definecolor{drawColor}{RGB}{231,41,138}

      \path[draw=drawColor,line width= 0.4pt,line join=round,line cap=round] (157.27,122.97) -- (162.82,122.97);

      \path[draw=drawColor,line width= 0.4pt,line join=round,line cap=round] (160.04,120.20) -- (160.04,125.75);
    \end{scope}
    \begin{scope}
      \definecolor{drawColor}{RGB}{231,41,138}

      \path[draw=drawColor,line width= 0.6pt,line join=round] (156.63,122.97) -- (163.46,122.97);
    \end{scope}
    \begin{scope}
      \definecolor{fillColor}{RGB}{255,255,255}

      \path[fill=fillColor] (204.08,118.70) rectangle (212.62,127.24);
    \end{scope}
    \begin{scope}
      \definecolor{drawColor}{RGB}{102,166,30}

      \path[draw=drawColor,line width= 0.4pt,line join=round,line cap=round] (206.39,121.01) rectangle (210.31,124.93);

      \path[draw=drawColor,line width= 0.4pt,line join=round,line cap=round] (206.39,121.01) -- (210.31,124.93);

      \path[draw=drawColor,line width= 0.4pt,line join=round,line cap=round] (206.39,124.93) -- (210.31,121.01);
    \end{scope}
    \begin{scope}
      \definecolor{drawColor}{RGB}{102,166,30}

      \path[draw=drawColor,line width= 0.6pt,line join=round] (204.93,122.97) -- (211.76,122.97);
    \end{scope}
    \begin{scope}
      \definecolor{drawColor}{RGB}{0,0,0}

      \node[text=drawColor,anchor=base west,inner sep=0pt, outer sep=0pt, scale=  0.80] at ( 52.38,120.22) {BIB};
    \end{scope}
    \begin{scope}
      \definecolor{drawColor}{RGB}{0,0,0}

      \node[text=drawColor,anchor=base west,inner sep=0pt, outer sep=0pt, scale=  0.80] at ( 87.14,120.22) {CIB};
    \end{scope}
    \begin{scope}
      \definecolor{drawColor}{RGB}{0,0,0}

      \node[text=drawColor,anchor=base west,inner sep=0pt, outer sep=0pt, scale=  0.80] at (122.00,120.22) {CX3/40};
    \end{scope}
    \begin{scope}
      \definecolor{drawColor}{RGB}{0,0,0}

      \node[text=drawColor,anchor=base west,inner sep=0pt, outer sep=0pt, scale=  0.80] at (170.31,120.22) {CX3/56};
    \end{scope}
    \begin{scope}
      \definecolor{drawColor}{RGB}{0,0,0}

      \node[text=drawColor,anchor=base west,inner sep=0pt, outer sep=0pt, scale=  0.80] at (218.62,120.22) {SoftRoCE};
    \end{scope}
  \end{tikzpicture}
\end{figure*}


\Cref{fig:mrs} shows the MR registration time, depending on the region's size.
MR registration costs are split between the OS and the NIC:
The OS pins the memory and the NIC learns about the virtual memory mapping of
the registered region.
SoftRoCE does not incur the \enquote{NIC-part} of the cost, therefore MR
registration with SoftRoCE is faster than for RDMA-enabled NICs.
For this experiment, we do not consider the costs of transferring the contents
of the MR during migration.

The number of QPs is the second variable influencing the migration time.
\Cref{fig:qps} shows the time for migrating a container running the \sendbw{}
benchmark.
The benchmark consists of two single-process containers running on two different
nodes.
Three seconds after the communication starts, the container runtime migrates one
of the containers to another node.
The migration time is measured as the maximum message latency as seen by the
container that did not move.
The checkpoint is transferred over the same network link used by the benchmarks
for communication.
%
%
With growing number of QPs, the benchmark consumes more memory, ranging from
\SI{8}{\mebi\byte} to \SI{20}{\mebi\byte}.
To put things into perspective, we estimated the migration time for real devices
by calculating the time to recreate \ibverbs{} object for RDMA-enabled NICs.
We subtracted the time to create \ibverbs{} objects with SoftRoCE from the
measured migration time and added time to create \ibverbs{} objects with
RDMA-NICs (from~\cref{fig:obj}).
We show our estimations with the dashed lines.





\subsection{MPI Application Migration}

\begin{figure}
  \centering
  \input{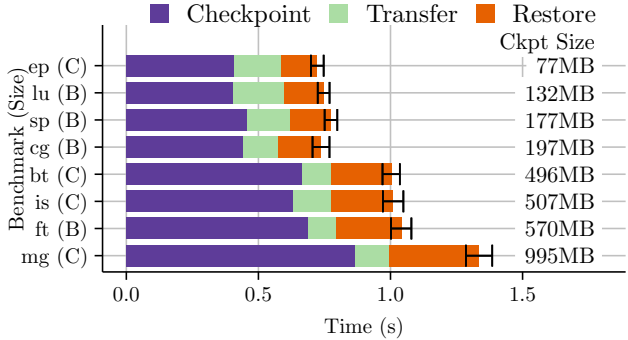}
  \caption[MPI migration]{MPI application migration.}
  \label{fig:mpi}
\end{figure}

For evaluating transparent live-migration of real-world applications we chose to
migrate NPB 3.4.1~\cite{baileyNasParallelBenchmarks1991}, an MPI benchmark
suite.
The MPI applications run on top of Open~MPI~4.0~\cite{openmpi}, which in turn
uses OpenUCX~1.6.1~\cite{ucx} for point to point communication.
We configured UCX to use \ibverbs{} communication over reliable connection (RC).

This setup corresponds to \cref{fig:sw-stacks}.
We containerised the applications using self-developed runtime CR-X, based on
libcontainer~\cite{runc}.
Unlike Docker, our container runtime facilitates faster live migration by
sending the image directly to the destination node, instead of the local
storage, during the checkpoint process.
Moreover, our container runtime stores checkpoint in RAM, reducing migration
latency even further.
The remaining description of our container runtime is out of scope of this
paper.

To measure the latency of application migration, we start each MPI application
with four processes (\emph{ranks}).
We migrate one of the ranks to another node approximately in the middle of the
application progress.
Each benchmark has a \emph{size} (A to F) parameter.
We chose size such that different benchmarks run between 10 and 300 seconds.
For this reason, we excluded dt benchmark, because it runs only around a second.
\Cref{fig:mpi} shows container migration latency and standard deviation around
the mean, averaged over 20 runs of each benchmark.

We break down the migration latency into three parts: \emph{checkpoint},
\emph{transfer}, and \emph{restore}.
\name{} stops the target container in the beginning of the checkpoint phase.
Large part of the checkpoint arrives to the destination node already during the
checkpoint phase.
After, the transfer phase is over, \name{} recovers the container on the
destination node.
Overall, we observe the migration time to be proportional to the checkpoint
size.
The benchmarks experience runtime delay proportional to the migration latency.

To show interoperability with other container runtimes, we measured migration
costs, when using Docker~19.03 (see~\cref{fig:docker}).
We had to implement full end-to-end migration flow ourselves, because Docker
supports only checkpoint and restore features.
To our disappointment, Docker does not employ some important optimisations and
takes line time to complete migration.
Nevertheless, we prove our claim that \name{} is readily interoperable with
other container runtimes.

%



\section{Related Work}
\label{sec:related}

\newcommand*{\y}{\ding{51}}
\newcommand*{\n}{\ding{55}}
\begin{table}
  \centering
  \begin{tabular}{lccccccc}
    \toprule{}
    & \rotatebox{90}{\footnotesize Legion}
    & \rotatebox{90}{\footnotesize Nomad}
    & \rotatebox{90}{\footnotesize PS~MPI}
    & \rotatebox{90}{\footnotesize DMTCP}
    & \rotatebox{90}{\footnotesize MOSIX-4}
    & \rotatebox{90}{\footnotesize MOSIX-3}
    & \rotatebox{90}{\footnotesize \name{}}\\
    \midrule
    RDMA         & \y & \y & \y & \y & \n & \n & \y \\
    Overhead     & N  & N  & N  & Y  & Y  & Y  & N  \\
    \midrule
    Runtime      & \y & \y & \y &    &    &    &    \\
    User-OS      &    &    &    & \y & \y &    &    \\
    Kernel-OS    &    & \y &    &    &    & \y & \y \\
    Hardware     &    &    &    &    &    &    & \y \\
    \midrule
    Units        & O  & VM & P  & P  & P  & P  & C \\%
    \midrule
    Reference    
    & \footnotesize \cite{legion}
    & \footnotesize \cite{nomad}
    & \footnotesize \cite{pickartzNonintrusiveMigrationMPI2016}
    & \footnotesize \cite{dmtcp}
    & \footnotesize \cite{barakMOSIXClusterManagement2016}
    & \footnotesize \cite{barakMOSIXDistributedOperating1993}
    & \footnotesize Ours\\
    \bottomrule
  \end{tabular}
  \caption{Selected checkpoint/restart systems handle either VMs, processes (P),
    containers (C), or application objects~(O).
    Runtime-based systems naturally introduce no additional communication
    overhead for migration support.
  }
  \label{tab:cr-cmp}
\end{table}

\paragraph{Checkpoint/Restart Techniques}
Transparent live migration of processes~\cite{smithSurveyProcessMigration1988,
  milojicicMobility1999, barakDistributedLoadbalancingPolicy1985},
containers~\cite{maEfficientServiceHandoff2017,
  mirkinContainersCheckpointingLive2008, nadgowdaVoyagerCompleteContainer2017}, or
virtual machines~\cite{nelsonFastTransparentMigration2005,
  clarkLiveMigrationVirtual2005, deshpandeFastServerDeprovisioning2014,
  hinesPostcopyLiveMigration2009, panCompSCLiveMigration2012} has long been a
topic of active research.
The key challenge of this technique lies in the checkpoint/restart operation.
For processes and containers, this operation can be implemented at three levels:
application runtime, user-level system, or kernel-level system.
\Cref{tab:cr-cmp} compares a selection of existing checkpoint/restart systems.

\emph{Runtime-based} systems expect the user application to access all external
resources through the API of the runtime system.
This restriction resolves two important issues with resource migratability:
First, the runtime system controls exactly when the underlying resource is used
and can easily stop the user application from doing so to serialise the state of
the resource.
Second, the runtime can maintain enough information about the state of the
resource to facilitate resource serialisation and deserialisation.
Such interception is cheap because it happens within the application's address
space.

Almost all attempts to provide transparent live migration together with RDMA
networks rely on modifications of the runtime
system~\cite{guayEarlyExperiencesLive2015, pickartzNonintrusiveMigrationMPI2016,
  dmtcp, nomad, joseSRIOVSupportVirtualization2013, gargMANAMPIMPIAgnostic2019}.
Some runtime systems operate on application-defined objects (tasks, agents,
lightweight threads) for even more efficient state serialisation and
deserialisation~\cite{legion, kaleCHARMPortableConcurrent1993,
  java-mobile-agents}.
All runtime-based approaches bind the application to a particular runtime
system.

%
%
%
%
%

\emph{Kernel OS-level} checkpoint/restart
systems~\cite{barakMOSIXDistributedOperating1993, blcr,
  osmanDesignImplementationZap2003, jakeedgeCheckpointRestartTries2009,
  kadavLiveMigrationDirectaccess2009} either do interposition at the kernel
level or extract application state from the kernel's internal data structures.
Although these systems support a wider spectrum of user
applications, they incur a significantly higher maintenance burden.
BLCR~\cite{blcr} has been abandoned eventually.
CRIU~\cite{criu}, currently the most successful OS-level tool for
checkpoint/restart, keeps necessary Linux kernel modification at a minimum and
does not require interposing user-kernel API.
We describe this tool in more detail in~\cref{sec:criu}.
%
%
%
%

%

Finally, \emph{user OS-level} systems interpose the user-kernel API, providing
the same transparency and generality as kernel-based implementation.
Such systems use the \lstinline[language=bash]{LD_PRELOAD} mechanism to
intercept system calls from applications and virtualise system resources, like
file descriptors, process IDs, and sockets.
In version 4, MOSIX has been redesigned to work entirely at the user
level~\cite{barakMOSIXClusterManagement2016}.
DMTCP~\cite{dmtcp} is a transparent fault-tolerance tool for distributed
applications with support for \ibverbs{}.
To be able to extract the state of \ibverbs{} objects, DMTCP maintains
\emph{shadow objects}, which act as proxies between a user process and the
NIC~\cite{caoTransparentCheckpointrestartInfiniband2014}.
In \cref{sec:eval-migratability}, we show that maintaining these shadow objects
has non-negligible runtime overhead for RDMA networks.

\paragraph{Network Virtualisation}

TCP/IP network virtualisation is an essential tool for isolating distributed
applications from the underlying physical network topology.
Even though network virtualisation enables live migration, it introduces
overhead due to additional encapsulation of network
packets~\cite{niuNetKernelMakingNetwork2019,slimos}.
%
%
Several new approaches try to address these performance problems~\cite{arrakis,
  ix, slimos, niuNetKernelMakingNetwork2019}.
However, these approaches do not consider RDMA networks.

Other work focuses on virtualising RDMA networking.
FreeFlow~\cite{freeflow} intercepts communication via \ibverbs{} in containers
to implement connection control policies in software but does not support live
container migration.
Nomad~\cite{nomad} uses \infiniband{} address virtualisation for VM migration
but implements the connection migration protocol inside an application-level
runtime.
LITE~\cite{tsaiLITEKernelRDMA2017} virtualises RDMA networks, but offers no
migration support and requires application rewrite.
%

%

%
%
%

\name{} uses traditional network virtualisation for TCP/IP networks, which is
not on the performance-critical path for RDMA-applications.
However, \name{} avoids unnecessary interception of RDMA-communication.
Instead, \name{} silently replaces addressing information during migration.

\paragraph{RDMA Implementations}

There are multiple open-source RDMA implementations.
SoftRoCE~\cite{liranlissLinuxSoftRoCEDriver2017} and SoftiWarp~\cite{softiwarp}
are pure software implementations of
RoCEv2~\cite{SupplementInfiniBandArchitecture2014} and iWarp~\cite{iwarp}
respectively.
Both provide no performance advantage over socket-based communication but
are compatible with their hardware counterparts and facilitate the development
and testing of RDMA-based applications.
We chose to base our work on SoftRoCE because RoCEv2 found wider adoption than
iWarp.

There are also open-source FPGA-based implementations of network stacks.
NetFPGA~\cite{zilbermanNetFPGARapidPrototyping2015} does not support RDMA
communication.
StRoM~\cite{sidlerStRoMSmartRemote2020} provides a proof-of-concept RoCEv2
implementation.
However, we found it unfit to run real-world applications (for
example, MPI) without further significant implementation efforts.


\section{Discussion}
\label{sec:discussion}

\paragraph{Hardware Modifications and Software Implementation}
Propositions to modify hardware often meet criticism because they tend to be
hard to validate in practice.
We believe that limited hardware changes are worthy of consideration as the
deployment of custom~\cite{kochevar-curetonRemoteDirectMemory2019, accelnet},
programmable~\cite{kochevar-curetonRemoteDirectMemory2019,
moonAccelTCPAcceleratingNetwork2020}, or software-augmented
NICs~\cite{hanSoftNICSoftwareNIC2015} has already been proven feasible.
\ibverbs{} has routinely been extended with additional
features~\cite{ibverbs-odp, xrc} as well.
Deploying \name{} to real data centres would require hardware changes.
We believe this trade-off is justified because \name{} provides tangible
performance benefits, in comparison to other approaches.

To find out whether our proposed changes have any effect on the critical path of
the communication, we integrated them into a software implementation of RoCEv2.
Our measurements show no performance difference after adding support for
migration.
Given the nature of these changes, we are confident this observation applies to
hardware as well.
Moreover, we provide our open-source software implementation to the research
community for validating our findings and further study.



\paragraph{Compatibility with Existing Infrastructure}
\name{} ensures by design backwards compatibility at the \ibverbs{} API and
RoCEv2 protocol level.
Moreover, \name{} allows to use container runtimes interchangeably.
By enabling migratability through \name{}, a data centre provider does not have
to make the hard choice of punishing applications that do not benefit from
migration.
We believe these features are crucial for successful integration into existing
data centre management infrastructure.

\paragraph{Unreliable Datagram Communication}
\name{} provides live migration for reliable communication~(RC), but omits
unreliable datagram~(UD) communication for two reasons:
First, every message received over UD exposes the address of its sender.
When this sender migrates, its address will change and currently \name{} cannot
conceal this fact from the receiver.
Second, a UD~QP can receive messages from anywhere.
This means that a UD~QP does not know where to the send resume messages after
migration.
We leave migration support for unreliable datagram for future work.

\section{Conclusion}

We introduce \name{}, an OS-level architecture enabling transparent live
container migration.
Our architecture design maintains full backwards-compatibility and
interoperability with the existing RDMA network infrastructure at every level.
We demonstrate end-to-end migration flow of MPI applications using different
container runtimes and studied cost of migration.
\name{} provides live migration without sacrificing RDMA network performance,
yet at the cost of changes to the RDMA communication protocol.

To validate our solution, we integrated the proposed RDMA communication protocol
changes into an open-source implementation of the RoCEv2 protocol, SoftRoCE.
For real-world deployment, these protocol changes must be implemented in NIC
hardware.
Finally, we provide a detailed analysis of any changes we make to SoftRoCE to
show their smallness.

We are convinced the architecture of \name{} can be useful for dynamic load
balancing, efficient prepared fail-over, and live software updates in data
centres or HPC clusters.



\section*{Acknowledgments}

The research and the work presented in this paper has been supported by the
German priority program 1648 “Software for Exascale Computing” via the research
project FFMK~\cite{ffmk-web}.
This work was supported in part by the German Research Foundation (DFG) within
the Collaborative Research Center HAEC and the the Center for Advancing
Electronics Dresden (cfaed).
The authors are grateful to the Centre for Information Services and High
Performance Computing (ZIH) TU Dresden for providing its facilities for high
throughput calculations.
In particular, we would like to thank Dr.~Ulf~Markwardt and Sebastian Schrader
for their support with the experimental setup.
The authors acknowledge support from the AWS Cloud Credits for Research for
providing cloud computing resources.


\section*{Availability}

The anonymised version of the code is available here:
\href{https://www.dropbox.com/s/clych73kxmuwjrt/asplos21-artifacts-125.tar.gz}{dropbox.com/s/clych73kxmuwjrt}.


\bibliographystyle{plainnat}
\bibliography{../zotero}

\end{document}